\documentclass[reprint, amsmath,amssymb, aps, showkeys, floatfix, nofootinbib]{revtex4-2}

\usepackage{layouts}
\usepackage{natbib}
\setcitestyle{authoryear,open={(},close={)}}
\usepackage{cancel}

\usepackage[dvipsnames]{xcolor}
\definecolor{mycitecolor}{HTML}{0071BC}
\definecolor{mybibcolor}{HTML}{15998E}
\usepackage[colorlinks=true,linkcolor=black,citecolor=mycitecolor, urlcolor=mybibcolor]{hyperref}%
\def\equationautorefname~#1\null{Equation (#1)\null}

\usepackage{graphicx}
\usepackage{dcolumn}
\usepackage{bm}
\usepackage{amsthm}
\usepackage{algorithmic}
\usepackage{centernot}
\usepackage[normalem]{ulem}
\usepackage{svg}
\usepackage{xprintlen}
\usepackage{adjustbox}
\usepackage{booktabs}

\usepackage{afterpage}
\newcommand\myemptypage{
    \null
    \thispagestyle{empty}
    \addtocounter{page}{-1}
    \newpage
    }
    
\providecommand{\keywords}[1]
{
  \small	
  \textbf{\textit{Keywords---}} #1
}

\newcommand{\dat}{\textbf{d}}
\newcommand{\parvec}{\boldsymbol{\theta}}

\newcommand{\Covar}{\textbf{C}}

\DeclareMathOperator{\tr}{tr}

\usepackage[normalem]{ulem}

\begin{document}

\preprint{APS/123-QED}

\title{Hybrid summary statistics: neural weak lensing inference beyond the power spectrum}

\author{T. Lucas Makinen}
\email{l.makinen21@imperial.ac.uk}
\affiliation{%
    Imperial Centre for Inference and Cosmology (ICIC) \& Imperial Astrophysics,
    Imperial College London, Blackett Laboratory, Prince Consort Road, London SW7 2AZ,
    United Kingdom
}%

\author{Alan Heavens}%
\affiliation{%
Imperial Centre for Inference and Cosmology (ICIC) \& Imperial Astrophysics,
Imperial College London, Blackett Laboratory, Prince Consort Road, London SW7 2AZ,
United Kingdom
}

\author{Natalia Porqueres}%
\affiliation{%
Department of Physics, University of Oxford, Denys Wilkinson Building, Keble Road, Oxford OX1 3RH, United Kingdom
}%

\author{Tom Charnock}
\noaffiliation

\author{Axel Lapel}
\affiliation{%
 Sorbonne Université, CNRS, UMR 7095, Institut d’Astrophysique de Paris, 98 bis boulevard Arago, 75014 Paris, France
}
\affiliation{Sorbonne Universit\'e, Universit\'e Paris Diderot, Sorbonne Paris Cit\'e, CNRS, Laboratoire de Physique Nucléaire et de Hautes Energies (LPNHE). 4 place Jussieu, F-75252, Paris Cedex 5, France}
 
\author{Benjamin D. Wandelt}
\affiliation{%
 Sorbonne Université, CNRS, UMR 7095, Institut d’Astrophysique de Paris, 98 bis boulevard Arago, 75014 Paris, France
}%
\affiliation{%
Center for Computational Astrophysics, Flatiron Institute, 162 5th Avenue, New York, NY 10010, USA
}%

\date{\today}

\begin{abstract}
    In inference problems, we often have domain knowledge which allows us to define summary statistics that capture most of the information content in a dataset.  In this paper, we present a hybrid approach, where such physics-based summaries are augmented by a set of 
    compressed neural summary statistics that are optimised 
    to extract the extra information that is not captured by the predefined summaries. The resulting statistics are very powerful inputs to simulation-based or implicit inference of model parameters. We apply this generalisation of Information Maximising Neural Networks (IMNNs) to parameter constraints from tomographic weak gravitational lensing convergence maps to find summary statistics that are explicitly optimised to complement angular power spectrum estimates. We study several dark matter simulation resolutions in low- and high-noise regimes. We show that i) the information-update formalism extracts at least $3\times$ and up to $8\times$ as much information as the angular power spectrum in all noise regimes, ii) the network summaries are highly complementary to existing 2-point summaries, and iii) our formalism allows for networks with smaller, physically-informed architectures to match much larger regression networks with far fewer simulations needed to obtain asymptotically optimal inference. 
\end{abstract}


\maketitle

\section{\label{sec:intro}Introduction}
Weak gravitational lensing alters the trajectories of distant photons as they pass through the large-scale structure of visible and dark matter to our detectors. These deflections alter the observed shapes of galaxies, whose patterns can be used to trace the matter distribution in between, and are sensitive to parameters that describe the expansion history and structure formation of the Universe. {The inference of these parameters from cosmological weak lensing surveys} is usually performed using two-point statistics of the lensing images, such as shear correlation functions or power spectra. Recent analyses include  the Dark Energy Survey \citep{amon_des_2021_22, secco_2022_des}, the Kilo-Degree Survey \citep[KiDS,][]{Li_kids, asgari_kids} and the Hyper Suprime-Cam survey \citep{dalal_hsc}. However, two-point statistics do not fully describe the rich non-Gaussian features present in large-scale structure, where more cosmological information might be found. 

Implicit inference (also known as simulation-based inference or likelihood-free inference) has made it possible to utilise higher-order statistics derived from simulations (see e.g. \cite{Cranmer30055} for a review), and circumvent the need for { an explicit likelihood function, which can be challenging to compute via Bayesian Hierarchical Models \citep{alsing_lensing_2017, porqueres_borg_2021, natalia2021, porqueres2023fieldlevel, almanac_23, almanac_23b}}. In weak gravitational lensing for example, even (incorrectly) assuming that the underlying cosmological density is Gaussian, the two-point statistics that describe this field can themselves have significantly non-Gaussian sampling distributions \citep{selletin_lensing_2017, alsing_lensing_2017, max_kids_sbi}. 
{The question arises as to which additional statistics contain significant extra information about the parameters, beyond that which is present in the two-point functions. Higher-order statistics are an obvious choice, but they suffer from a lack of knowledge of their sampling distributions, and the very large number of statistics make them cumbersome for implicit inference. However,} advances in deep learning have made it possible to learn {highly-informative} neural compressions for massive simulations automatically,  and in some cases losslessly \citep{makinen2021, Charnock_IMNN, Makinen_2022_graphs}. {These compressions yield radically smaller summary spaces, which {are ideal for implicit inference, and which} can be used for Bayesian posterior estimation via accept-reject or density estimation strategies \citep{pydelfiAlsing_2019}. }

These new advances have made simulation-based studies for weak lensing a popular avenue of research in recent years. This includes studies of peak counts \citep{peakcounts_lensing_constraints, DESpeaks, peakcounts_lensing2010, Lanzieri_2023_higherorder}, higher-order statistics \citep{ajani_et_al_euclid} such as wavelet scattering transformations \citep{cheng_scattering, scattering_cheng2024cosmological}, Fourier-space normalising flows \citep{dai2024multiscale}, and field-based convolutional neural networks \citep{zoltan_Ribli_2019, Fluri_2018_convergence, Fluri_2019lensing, sharma2024comparative}. The ultimate goal is to obtain statistics that exhaust the information content of the observed weak lensing field. This can be done explicitly (assuming a likelihood for shear or convergence voxels) via field-level sampling \citep{alsing_lensing_2017,2019A&A...621A..69R, natalia2021, porqueres_borg_2021, porqueres2023fieldlevel, florent_borg, Jasche_2015, boruah2023mapbased_supranta, almanac_23, tsaprazi2023}. 

Impicit inference approaches have matured enough for real-data analysis, beginning with \cite{fluri_kids}, who analysed map-level KiDS data with an assumed Gaussian summary likelihood, and more recently \cite{max_kids_sbi}, who reproduced $C_\ell$ constraints with an implicit likelihood. \cite{jeffrey2024dark} presented a Dark Energy Survey data analysis using a convolutional neural network compression of the full mass map and demonstrated marked improvement over existing power spectrum and peak count constraints on the same dataset. 

This work seeks to demonstrate an improved optimisation strategy and add a new layer of interpretability to this growing body of literature. A common question for deep learning and implicit inference practitioners is what features are being learned from the data by neural approaches. \cite{Makinen_2022_graphs} showed explicitly that neural networks trained on halo catalogues identified features that could be linked to explicitly-understood cosmological distributions such as halo mass and correlation functions. Here we respond to this question by modifying our optimisation criterion such that a network only outputs statistics obtained from the data that work alongside to an existing statistic, in this case the weak lensing angular power spectrum. {To be explicit, we train the network to maximise the extra Fisher information that is not already present in the power spectrum.} We term these neural summaries ``hybrid'' statistics since they combine new and existing functions of the data.  We make our constraint comparison within a completely simulation-based setup to interrogate the information content of the respective summaries in a sampling distribution-agnostic way.

This paper is organised as follows: In Sections \ref{sec:sbi} and \ref{sec:formalism} we present our general formalism for finding optimal new summaries from simulated data given some existing descriptive statistics, and describe how the strategy can be implemented automatically with Information Maximising Neural Networks \citep[IMNNs,][]{Charnock_IMNN, makinen2021}. In Section \ref{sec:weaklensing}, we describe our mock weak lensing formalism and present the simulation suite details. In Section \ref{sec:wl_stats}, we present our angular $C_\ell$ compression scheme as the existing statistic in the information-update formalism. We then present our physics-inspired, lightweight neural network architecture designed to find optimal additional summaries. In Section \ref{sec:results}, we make comparisons of information gain over the power spectrum as a function of resolution and increased shape noise, and show that our optimisation scheme captures physical features in realistic noise regimes. 

\section{Implicit Inference}\label{sec:sbi}
The goal of most science experiments is to obtain data $\dat$ with which to test models that describe the data generation process. In cosmology, this often boils down to obtaining a posterior distribution for some model parameters $\parvec$: $p(\parvec | \dat) \propto p(\dat | \parvec) p(\parvec)$, which requires knowledge of the likelihood $p(\dat | \parvec)$, and the assumption of a suitable prior $p(\parvec)$. This data-generating distribution is often too complicated to evaluate for inference, or too complex to write down analytically.

\subsection{Density Estimation}
Simulation-based inference circumvents the need for a tractable likelihood $p(\dat | \parvec)$, and instead seeks to parameterise the underlying, implicit likelihood or posterior present in forward simulations of the data. Neural density estimators \citep[NDEs; e.g.][]{bishop_mdn_1994} use neural networks that give some estimate $q(\parvec, \dat; \varphi )$ of the desired posterior (or likelihood) by varying weights and biases (parameterised as $\varphi$) to minimize the loss
\begin{equation}\label{eq:nde_loss}
    U(\varphi) = - \sum_{i=1}^{N} \ln q(\parvec_i | \dat_i; \varphi),
\end{equation}
over batches of parameter-data samples drawn from the joint distribution ($\parvec_i, \dat_i) \curvearrowleft p(\parvec, \dat_i)$. This loss is equivalent to minimising the Kullback-Leibler divergence {between the target distribution and $q$ }\citep{kl_1951}. In this work we employ Masked Autoregressive Flows  \citep[MAF;][]{papa_mafs2017} to model the posterior distribution directly. We detail our implementation in Section \ref{sec:nde}. Density estimation also makes posterior predictive and coverage tests far easier to perform. We show examples of these tests in Appendix \ref{app:coverage}.

\subsection{Data Compression}
Accept-reject and density estimation schemes become more difficult to compare to a target, observed data vector $\dat_{\rm obs}$ the larger the dimensionality $\dim(\dat)$, which posits the need for data compression to some smaller summary space $\textbf{x}$. We would like a function $f: \dat \mapsto \textbf{x}$ that is ideally maximally informative about the parameters $\parvec$. Under certain conditions, $f(\textbf{d})$ can yield a sufficient statistic, for which the dimension $\textbf{x}$ is equal to the number of parameters, e.g. $\dim(\textbf{x}) = \dim(\parvec)$. \cite{Heavens_2000} introduced Massively Optimised Parameter Estimation and Data (MOPED) compression, which gives optimal score compression for cases where the likelihood and sampling distributions are Gaussian, and this was generalised to other forms of score compression by \cite{alsing2018_general, CarronSzapudi, hoffmann2023minimising}. Neural compression is a popular scheme for learning mappings agnostic to sampling distributions, for which several optimisation schemes have been proposed.  Regression-style approaches learn a compression $f(\textbf{d}; w)$ parameterised by (neural) weights $w$ via a loss, for example quadratic, over parameter-data pairs over a prior, using variants of the mean square error (MSE), or in some cases learning $f$ and the neural posterior (via Eq. \ref{eq:nde_loss}) simultaneously, dubbed Variational Mutual Information Maximisation \citep[VMIM;][]{Jeffrey_massmap_vmim}. \cite{sharma2024comparative} recently compared these losses paired with convolutional networks for a separate weak lensing simulation suite.

This work builds upon the Information Maximising Neural Network (IMNN) approach \citep{Charnock_IMNN, makinen2021}, which prescribes a neural compression that maximises the determinant of the Fisher matrix of summaries around a \textit{local} point in parameter space. A compression is i) learned at a fiducial point from a set of dedicated fiducial and derivative simulations and then ii) applied to simulations over a prior for posterior construction. This approach has numerous advantages, namely:
\begin{enumerate}
    \item An asymptotically-optimal compression can be learned from simulations around a single point in parameter space.
    \item The compression automatically {and simultaneously} gives Fisher posterior constraint forecasts.
    \item Priors used for density estimation {are decoupled from the compression step and} can be chosen after learning the compression.
    \item Adding additional parameters of interest to the compression learning scheme only requires relatively small numbers of extra simulations {for the new derivatives}; e.g. the distribution $p(\parvec, \dat)$ need not be re-simulated. 
\end{enumerate}
In the following section we will extend this approach to find new (neural) data compressions that only increase information about parameters {above what is already present in a set of existing statistics such as the power spectrum}.

\section{How to Choose an Optimal New Summary}\label{sec:formalism}
Consider some data $\textbf{d} \in \mathbb{R}^N$ created from parameters $\boldsymbol{\theta}$ that can be summarised in a compressed summary vector via a function $h: \dat \mapsto \textbf{t}$ with $\textbf{t} \in \mathbb{R}^{n_t}$ where $n_t < N$. We can estimate the covariance matrix of the summaries  $\textbf{C}_t$, and the mean $\boldsymbol{\mu}_t$ from simulations, along with derivatives with respect to parameters of the mean $\boldsymbol{\mu}_{,\theta_i}$. {Assuming for now that the summaries have a Gaussian sampling distribution (this assumption is temporary and is dropped in the inference phase)}, we can compute the Fisher information of the observables via
\begin{equation}\label{eq:fisher_eq}
    [\textbf{F}_t]_{ij} = \boldsymbol{\mu}_{,\theta_i}^T \Covar^{-1}_t \boldsymbol{\mu}_{,\theta_j}
\end{equation}
where we introduce the notation {$\bm{y}_{,\theta_i}\equiv{\partial \bm{y}}/{\partial\theta_i}$} for partial derivatives
with respect to parameters. The Fisher information matrix here
describes how much information $h(\dat)$ contains about the model parameters, and is given as the second moment of the score of the
likelihood with respect to $h$, assuming a parameter-independent, Gaussian covariance of the statistic $\textbf{t}$.\footnote{{Note that the Gaussian assumption is used here only to define a compression; once the summaries are defined, SBI no longer assumes Gaussianity.  If the compressed summaries are not Gaussian-distributed, the compression will be suboptimal but the downstream SBI analysis will implicitly determine and use their true sampling distribution.}} A large Fisher information for a function of the data indicates that the {mapping to $\textbf{t}=h(\textbf{d})$} is very informative about the model parameters used to generate the realisation of data $\dat$. Fisher forecasting for a given model is made possible by the information inequality and the Cram\'er-Rao bound \citep{cramerharald_1946, rao_1945}, which states that the minimum variance of the value of an estimator $\parvec$ is given by 
\begin{equation}\label{eq:cramer-rao}
    \langle (\theta_i - \langle \theta_i \rangle )^2  \rangle \geq (\textbf{F}^{-1})_{i i},
\end{equation}
{with no summation over $i$.}

\begin{figure*}[htp!]
    \centering
    \includegraphics[width=0.95\textwidth]{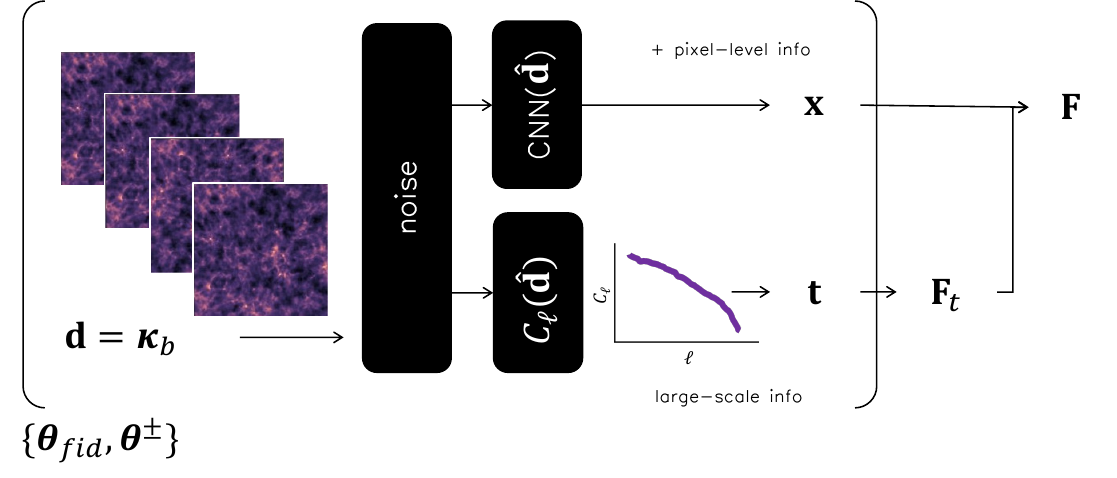}
    \caption{Hybrid summary network schematic, illustrated for weak gravitational lensing. Noisy data (weak lensing $\boldsymbol{\kappa}_b$ with shape noise) are passed in parallel to an existing summary function (tomographic $C_\ell$ with optional MOPED compression) to produce summaries $\textbf{t}$, and a network (CNN) to output an additional set of summaries $\textbf{x}$, described in Section \ref{sec:wl_stats} and illustrated in Figure \ref{fig:neuralnet}. To train the network the Fisher information is first calculated for $\textbf{t}$ and then updated via Equation \ref{eq:matrixupdate} to yield $\textbf{F}$, for the loss Eq. \ref{eq:imnn-loss}. }
    \label{fig:imnn-diagram}
\end{figure*}

{We would next like to add new summary statistics to increase the information over what is present in $\textbf{t}$.  For clarity, we begin by adding a single summary ${x}$, before generalising to multiple additional summaries.} We do this via some function $f: \textbf{d} \mapsto {x}$. This new number has variance $\sigma^2_x$ and when concatenated to the old observables gives the mean vector
\begin{equation}
    \boldsymbol{\mu} = [\boldsymbol{\mu}_t, \langle x \rangle]^T.
\end{equation}
For mean-subtracted quantities $\Delta t$ and $\Delta x$, the covariance vectors between old and new observables can be computed (e.g. from simulations) as $[\textbf{u}]_i = \langle  \Delta t_i \Delta x \rangle$, which yields the full covariance matrix
\begin{equation}\label{eq:fullcov}
    \textbf{C} = \begin{pmatrix}
    \textbf{C}_t & \textbf{u}\\
    \textbf{u}^T & \sigma^2_x
\end{pmatrix}.
\end{equation}
Notice here that the smaller the values of $\textbf{u}$, the less correlated $x$ is with $\textbf{t}$. The full updated Fisher matrix is then
\begin{equation}\label{eq:bruteforceFisher}
    F_{ij} = \boldsymbol{\mu}^T_{,\theta_i} \Covar^{-1} \boldsymbol{\mu}_{,\theta_j}.
\end{equation}
With some rearrangement, we obtain the fast Information-Update Formula (IUF):
\begin{equation}\label{eq:IUF}
    \textbf{F} = \textbf{F}_t + \frac{1}{s} \textbf{v}\textbf{v}^T,
\end{equation}
with $[\textbf{v}]_i = \langle x \rangle_{,\theta_i} - \boldsymbol{\mu}^T_{,\theta_i} \textbf{C}_t^{-1} \textbf{u} $ and $s = \sigma_x^2 - \textbf{u}^T \textbf{C}_t^{-1} \textbf{u}$. This calculation is only $\mathcal{O}(n_{\rm params}^2 + n_{\rm params}d + d^2)$ operations where $d=\dim(\textbf{t})$, which is asymptotically $d$ times faster than Eq. \ref{eq:bruteforceFisher} when $d >> n_{\rm params}$.  This formalism also yields a fast update for the determinant of the new Fisher matrix:
\begin{equation}\label{eq:detF}
\ln \det \textbf{F} = \ln \det \textbf{F}_t + \ln \left( 1 + \frac{1}{s} \textbf{v}^T \textbf{F}^{-1}_t \textbf{v}  \right).
\end{equation}
\newline
\noindent\textbf{Interpretation.} The updated Fisher information in Equation \ref{eq:IUF} clearly separates the information contribution from the existing observables in the first Fisher term and the new observables in the second term. An optimal, ``complementary'' new observable $x$ adds a lot of information if it has highly correlated measurement error with the existing summaries $\textbf{t}$, but changes with respect to parameters in a way that is as distinguishable as possible from how $x$ and $\textbf{t}$ change together.
\newline




\noindent \textbf{Multiple New Observables.} The IUF can be naturally extended to a vector of new summaries $\textbf{x}$. We promote $\textbf{v}$ to a matrix
\begin{equation}
    \left[ \boldsymbol{V} \right]_{ij} = \langle {\textbf{x}_j} \rangle_{,\theta_i} - \boldsymbol{\mu}^T_{,\theta_i} \textbf{C}_t^{-1} \textbf{u}_j,
\end{equation}
and the scalar $s$ generalises to the matrix 
\begin{equation}
    \boldsymbol{\Sigma} = \Covar_{x} - \textbf{U}^T \Covar^{-1}_t \textbf{U}
\end{equation}
where $[\textbf{U}]_{ij} = [\textbf{u}_j]_i$ and $\Covar_x$ is the covariance of network outputs $\textbf{x}$. Altogether the updated Fisher matrix for a vector of extended summaries is
\begin{equation}\label{eq:matrixupdate}
    \textbf{F} = \textbf{F}_t + \boldsymbol{V} \boldsymbol{\Sigma}^{-1} \boldsymbol{V}^T.
\end{equation}

\subsection{Finding a New Summary With a Neural Network}
We can find a new observable $\textbf{x}$ by optimising the IUF equation (\ref{eq:IUF}) with a neural network $f: \textbf{d} \mapsto \textbf{x}$ that operates on the data. This formalism folds neatly into the IMNN formalism \citep{Charnock_IMNN, makinen2021, Makinen_2022_graphs}. We illustrate this procedure for weak lensing data in a schematic in Figure \ref{fig:imnn-diagram}.  We can choose to optimise Equation \ref{eq:bruteforceFisher} directly, but Equation \ref{eq:matrixupdate} is less computationally expensive for large covariance matrices and more than one additional summary. The ingredients needed to compute the components of the loss are a suite of $n_s$ simulations at a fiducial value of parameters $\{ \textbf{d} \}_{i=1}^{n_s}|_{\parvec = \parvec_{\rm fid}}$ and a set of $n_d$ seed-matched simulations at perturbed values of each parameter $\parvec_i^\pm = \parvec_i \pm \Delta \parvec_i$ holding all other parameters fixed at their fiducial values. Using this finite difference gradient dataset the partial derivatives of a data summary function $Q(\dat)$ with respect to parameters is
\begin{equation}\label{eq:num-deriv}
    \left( \frac{\partial \hat\mu_i}{\partial \theta_\alpha} \right) \approx \frac{1}{n_d}\sum^{n_d}_{i=1} \frac{Q(\dat^{+}_i) - Q(\dat^{-}_i)}{ \theta^+_\alpha - \theta^-_\alpha}.
\end{equation}
For $n_{\rm params}$ summaries, this method requires $n_d \times n_{\rm params} \times 2$ simulations with $n_d$ unique random seeds alongside the $n_s$ simulations at the fiducial point required for the covariance. This is done for the mean of both existing and new summary statistics, consolidated as $\textbf{{y}} = [\textbf{t}, \textbf{x}]$. The covariance of
the (existing and new) summaries is {estimated} from the data as well, using $n_s$ simulations at $\parvec_{\rm fid}$:
\begin{equation}\label{eq:imnn-cov}
    \hat{\textbf{C}}_{\alpha \beta} = \frac{1}{n_s - 1}\sum_{i=1}^{n_s} (\textbf{y}_i - \bar{\textbf{y}})_\alpha (\textbf{y}_i - \bar{\textbf{y}})_\beta,
\end{equation}
where $\bar{\textbf{y}}$ is the average over the simulations at the fiducial point.
The full covariance can be broken down into (or estimated separately by) its components $\textbf{C}_t, \textbf{C}_x$, and $\textbf{u}$ according to Eq. \ref{eq:fullcov}. Note that this covariance is assumed to be independent of
the parameters, which, whilst not strictly true, is
enforced by regularisation during the fitting of a network. {If it does not hold, it simply makes the compression suboptimal elsewhere in the parameter space.}
Crucially, both old and new summaries and their statistics must be computed on the same (noisy) simulations to correctly distinguish between noise fluctuations and newly-informative features of the data during optimisation.

Optimising a neural function $\textbf{x} = f(\dat ; w)$ to maximise the determinant of $\textbf{F}$ from Eq. \ref{eq:matrixupdate} forces the new summaries $\textbf{x}$ to add complementary information to the existing summaries' Fisher contributions. 
As described in \citet{Charnock_IMNN} and \citet{livet2021catalogfree}, the Fisher information is invariant to nonsingular linear transformations of the summaries. To remove this ambiguity, a term penalising the network summary covariance $\Covar_x$ is added. { This gives the loss function}
\begin{equation}\label{eq:imnn-loss}
\Lambda = -\ln \det \textbf{F}  + \lambda \frac{1}{2} \tr \Covar_x 
\end{equation}
where $\lambda$ is a regularising coefficient. This scalar loss function can be optimised via gradient descent to update weights $w$ for the network's contributions to the combined Fisher information. The network can do no worse at summarising the data than the existing summary, since the loss only optimises the second term of Eq. \ref{eq:matrixupdate}. With the updated Fisher information we can also compute quasi-maximum likelihood estimates (MLE) for the parameters for a given mean-subtracted summary vector $\Delta = [\Delta \textbf{t}, \Delta \textbf{x}]$ \citep{alsing2018_general}: 
\begin{equation}\label{eq:mle-from-fisher}
    \hat{\parvec} = \parvec_{\rm fid} + \textbf{F}^{-1} \boldsymbol{\mu}_{, \theta_i} \textbf{C}^{-1} \Delta^T.
\end{equation}
{We then use these hybrid statistics (as they are functions of the data) as our highly-informative and extremely compressed data set, ideal for simulation-based or implicit inference.}
The resulting compression is \textit{asymptotically} optimal at the fiducial point in parameter space, but for smoothly-varying data manifolds results in a smooth summary space that can be exploited for neural posterior estimation away from the fiducial point as described in \cite{makinen2021, Charnock_IMNN}. A useful aspect of learning this local compression is that a prior for posterior estimation can be specified after the compression network is trained, unlike regression networks.

\section{Weak Gravitational Lensing}\label{sec:weaklensing}

\subsection{Formalism}
The effect of weak lensing (WL) on a source field is defined by its shear, $\gamma$, which captures the distortions in the shapes of observed galaxies. In the flat-sky limit in Fourier space, this observable can be related to the convergence field $\kappa$ observable, which describes variation in angular size:
\begin{equation}\label{eq:shear}
    \tilde{\gamma}(\boldsymbol{\ell}) = \frac{({\ell}_1 + i {\ell}_2)^2}{\ell^2} \tilde{\kappa}(\boldsymbol{\ell})
\end{equation}
where $\boldsymbol{\ell} = (\ell_1, \ell_2)$ is the complex wavevector. The convergence field can be connected to the underlying dark matter field by integrating the fractional overdensity along the line-of-sight to give \citep{kilbinger-wl-dm}:
\begin{equation}\label{eq:kappa_dm}
    \kappa(\boldsymbol{\vartheta} ) = \frac{3 H_0^2 \Omega_m}{2 c^2} \int_0^{r_{\rm lim}} \frac{r dr}{a(r)} g(r) \delta^f (r\boldsymbol{\vartheta}, r),
\end{equation}
where $\boldsymbol{\vartheta}$ denotes the coordinate on the sky, $r$ is the comoving distance, $r_{\rm lim}$ is the galaxy survey's maximum comoving distance, $\delta^f$ is the dark matter overdensity field at scale factor $a$, and, for a flat Universe
\begin{equation}
    g(r) = \int_r^{r_{\rm lim}} dr' n(r') \frac{r - r'}{r},
\end{equation}
is the integration of the redshift distribution $n(r)$ in the given comoving shell. In real-data analyses the data will be the cosmic shear, but here we restrict our analysis to noisy convergence maps. The forward model to generate $\boldsymbol{\kappa}$ consists of a cosmological parameter draw, $\parvec$, which is used to generate primordial fluctuations, $\delta^{\rm ic}$. Here the initial conditions are a Gaussian random field governed by the \cite{Eisenstein_1999} cosmological power spectrum $P(k; \theta)$, which includes baryonic acoustic oscillations (BAO). The cosmic initial conditions are then evolved forward via a specified non-linear gravity model $G(\delta^{\rm ic})$, which describes the growth of the large-scale structure (LSS). The evolved dark matter field $\delta^f$ is then used to generate the convergence field. Using the Born Approximation, we implement a discrete version of Equation \ref{eq:kappa_dm} using a summation over voxels to approximate the radial line-of-sight integrals:
\begin{equation}\label{eq:discrete-kappa-dm}
    \kappa^b_{mn} = \frac{3 H_0^2 \Omega_m}{2 c^2} \sum_{j=0}^N \delta^f_{mnj} \left[ \sum_{s=j}^{N} \frac{(r_s - r_j)}{r_s} n^b(r_s) \Delta r_s \right] \frac{r_j \Delta r_j}{a_j},
\end{equation}
where $b$ indexes the tomographic bin, and $m,n$ index the spatial pixels on the sky. The index $j$ indicates the voxel along the line-of-sight at the comoving distance $r_j$. The total number of voxels along the line-of-sight, $N$, is obtained from a ray tracer. $\Delta r_j$ is the length of the line segment inside voxel $j$, and $\delta^f_{mnj}$ is the discretized dark matter overdensity field. The comoving radial distance $r_s$ is the distance to the source. Each tomographic bin has a source redshift distribution $n^b(z_s)$. Once $\kappa^b_{mn}$ is computed, the convergence field $\dat = \{\hat{\kappa}^b_{mn}\}$ is obtained by adding uncertainties equivalent to the shape noise (and measurement error) in the shear field.  This is captured by zero-centred Gaussian white noise added pixel-wise with variance
\begin{equation}
    \sigma_n^2 = \sigma_\epsilon^2 \frac{ N_{\rm tomo}}{n_{\rm gal} A_{\rm pixel}^b},
\end{equation}
where $\sigma_\epsilon^2$ is the total galaxy intrinsic ellipticity dispersion, ${n_{\rm gal}}$ is the source galaxy density on the sky, and $A_{\rm pixel}^b$ is the angular size of the pixel in each tomographic bin. For Stage-IV weak lensing surveys like Euclid   $n_{\rm gal}$ will be $\sim 30\ {\rm arcmin}^{-2}$ and $\sigma_\epsilon \simeq 0.3$ \citep{Euclidcollab2021}. For network training purposes we introduce an amplitude scaling parameter $\sigma_n' = A\sigma_n$ that we report in terms of effective  source galaxy density.

\subsection{Simulation Details} 
We analyse several simulation suites at different resolutions to conduct our experiments. In all cases our physical box size is kept fixed at $L_x = L_y = 250\ {\rm Mpc}\ h^{-1}$ and $L_z = 4000\ {\rm Mpc}\ h^{-1}$ in comoving units, in a pixel grid of shape $(N_x, N_y, N_z) = (N, N, 512)$, where we vary $N$ to probe changing gravity solver scales. We utilise \texttt{pmwd} particle mesh (PM) simulations \citep{li2022pmwd} integrated for 63 timesteps to generate the nonlinear dark matter overdensity field for $N=[64,128]$ resolution and 100 timesteps for $N=192$ resolution. This controls the particle spacings $L/N$ which probe increasingly nonlinear scales described by the particle-mesh (PM) simulations. We compute the line-of-sight integral in comoving units before binning the $L_z$ dimension in redshift bins converted to comoving units via the cosmology-dependent change of variable. For this analysis we do not include lightcone effects. We choose our four tomographic redshift bins to be Gaussian, centred at $z = [0.5, 0.75, 1.0, 1.25]$ with width $\sigma_z = 0.14$, following \cite{porqueres_borg_2021}. The resulting convergence fields span a $3.58 \times 3.58 \ {\rm deg}^2$ field of view. Shape noise is added to the noise-free simulations before computing two-point or network statistics as described below. We generate two distinct datasets to i) construct a locally optimal compression and ii) perform posterior density estimation.

\noindent \textbf{Compression Simulations.} For a given resolution we generate two equally-sized datasets for training and validation of our network compression. To calculate network and two-point covariances described below we simulate $n_s=1500$ simulations at a fiducial cosmology $\parvec_{\rm fid} = (\Omega_m, S_8) = (0.3, 0.8)$. For finite-difference derivatives we simulate $n_d \times 2 \times n_p = 375 \times 2 \times 2$ seed-matched simulations at a perturbed parameter set $\parvec \pm \Delta \parvec_{\rm fid} = \parvec_{\rm fid} \pm (0.0115, 0.01)$. All other cosmological parameters were held fixed at Planck 2018 parameters \citep{Planck2018}. The total number of simulations used for optimal compression is thus 4500.
\newline
\noindent \textbf{Density Estimation Simulations.} Because our compression from the convergence field data is learned locally, we are free to choose our prior guided by the compression method's Fisher forecast. We simulate 5000 simulations over a uniform prior in $(\Omega_m, S_8)$, whose width is chosen according to the strategy described in Section \ref{sec:nde}. \newline

\section{Finding Hybrid Weak Lensing Statistics}\label{sec:wl_stats}

The information-update formula is perfectly suited to improving weak lensing $\Omega_m-\sigma_8$ parameter constraints with neural summaries. We would like to know if more cosmological parameter information can be extracted from the convergence field beyond a simulation-based tomographic angular $C_\ell$ statistic analysis, and in what resolution regimes. We present the MOPED scheme for angular $C_\ell$ compression and information-update neural network architecture.

\subsection{MOPED Angular $C_\ell$ Compression}

Here our existing summaries are either binned angular $C_\ell$ or MOPED-compressed vectors $\textbf{t}$ (without the optional Gram-Schmidt orthogonalisation employed in \cite{Heavens_2000}). We outline our setup below and display a schematic of the architecture in Figure \ref{fig:imnn-diagram}.

We compute empirical auto- and cross-spectra $C_\ell$ across the four noisy tomographic bins, resulting in 10 $C_\ell$ vectors. To test scale-dependent information, we optionally apply a maximum $\ell_{\rm cut}$ to each vector to mimic existing survey analyses. {To reduce the number of simulations needed to estimate the covariance matrix,} we bin each spectrum into six evenly-spaced $\ell$ bins weighted by $C_\ell$ value. We can estimate the covariance of the $C_\ell$ vector using the $n_s$ fiducial simulations, and the finite difference derivatives with respect to each parameter via Eq. \ref{eq:num-deriv}. Together, the Fisher matrix for these summaries is computed with Eq. \ref{eq:fisher_eq}. With these ingredients we can then perform the MOPED compression from mean-subtracted vectors {evaluated at a fiducial set of parameters} $\Delta = \left( \hat{C}_\ell - \langle{C_\ell}\rangle_{\rm fid} \right)$ down to score summaries $\textbf{t}$:
\begin{equation}\label{eq:moped-compress}
    \textbf{t} = \parvec_{\rm fid} + \boldsymbol{\mu}_{, \theta_i} \textbf{C}_t^{-1} \Delta^T.
\end{equation}
which can then be rescaled by the Fisher matrix to obtain an MLE of the parameters \citep{alsing2018_general}:
\begin{equation}
    \hat{\parvec}_{\rm MOPED} = \parvec_{\rm fid} + \textbf{F}_t^{-1} \textbf{t}.
\end{equation}
In practice, we replace $\textbf{t}$ with  $\hat{\parvec}_{\rm MOPED}$ as our existing MOPED statistics, which has covariance $\textbf{C}_t = \textbf{F}_t^{-1}$. These compressed summaries are the default $C_\ell$-based summaries that we feed into the normalising flow posterior estimation scheme (Section \ref{sec:nde}), as normalising flows are not guaranteed to work well with large inputs e.g. the 60-dimensional binned $C_\ell$ vector \citep{Cranmer30055}. For network optimisation however, the longer, binned power spectrum vector can be used to find $\hat{\parvec}_{\rm network}$. Changes in the $\ell$ bins with respect to noise and parameters increases the number of cross-correlations ($\textbf{u}$) with network summaries and encourage improved information extraction. We explore this option for training in noisy settings in Section \ref{sec:highnoise}. Both choices of statistics fit neatly into the existing statistic formalism described in Section \ref{sec:formalism}. 

\begin{figure}[htpb!]
    \centering
    \includegraphics[width=0.9\columnwidth]{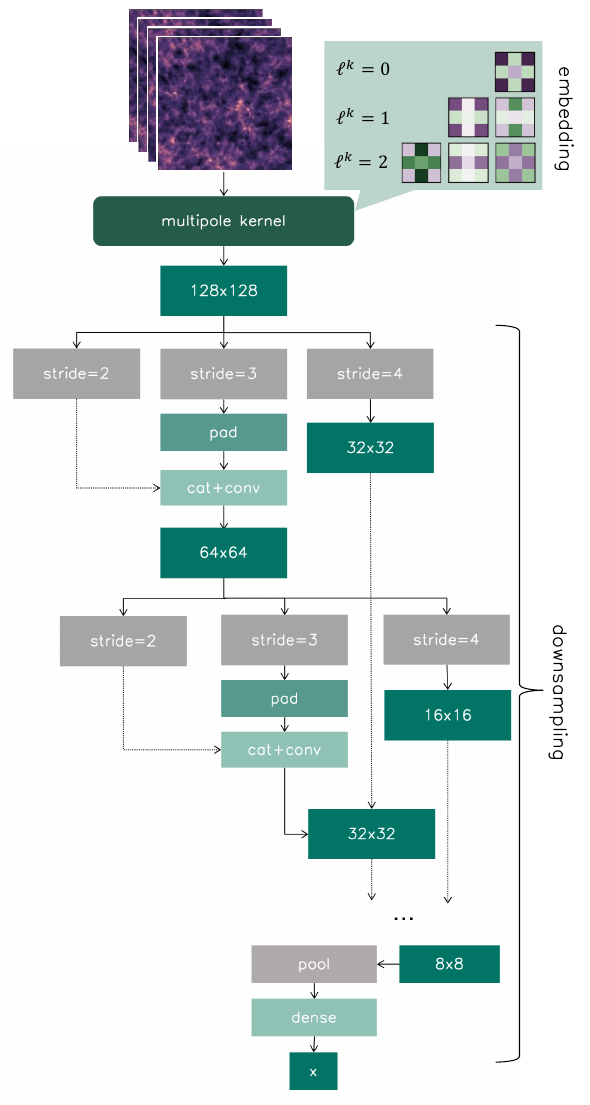}
    \caption{We use a small convolutional neural network that exploits the data symmetries to compress $\kappa$ fields down to additional summaries. Input data (here of shape $(128,128,4)$) are passed through a residual multipole kernel layer (shared colour indicates shared weights) and then subsequently passed to convolutional blocks with varying strides with small $2\times2$ kernels to capture fluctuations on different scales. All linear layers are followed by a nonlinear activation function. Dashed arrows indicate feature concatenation at the same spatial resolution. This downsampling continues until the spatial resolution of the data reaches $8\times 8$, after which the output tensor is mean-pooled along the spatial axes and passed to three dense layers. The output from the network is a pair of numbers.}
    \label{fig:neuralnet}
\end{figure}

\subsection{Physically-Informed Neural Network Architecture}
We design a 2D convolutional neural network with a dedicated structure to extract information in the convergence field data in a way that we know the information is likely to be distributed, and then learn a downsampling compression of these relevant features. We present this network layer-by-layer and display a schematic in Figure \ref{fig:neuralnet}. 

The inputs to the network are the log-transformation of the convergence maps at the four specified tomographic bins, adapted from \cite{Simpson_2015, Seo_2011, joachimi_2011}:
\begin{equation}
    \boldsymbol{\kappa}^b = \kappa_o \ln \left[ 1 + \boldsymbol{\kappa}^b / \kappa_o  \right]
\end{equation}
where $\kappa_o = |\kappa^b_{\rm min}| + 0.01$, where $\kappa^b_{\rm min}$ is the minimum convergence value for the tomographic bin $b$.
\newline

 \noindent \textbf{Multipole Kernel Embedding.} 
 For convergence maps we can target clustering information by learning convolution functions of the data with certain, enforced symmetries. \cite{Kodi_Ramanah_2020_superres} demonstrated via neural emulation of dark matter simulations that learned convolutional kernel weights tend to be distributed in spherically-symmetric ways. This motivates ordering kernel complexity by increasing breaking of rotational symmetry. \cite{Charnock_2020} and \cite{ding2024pinetreegenerativefastdifferentiable} explicitly encoded multipole expansion symmetries in CNN kernels to learn bias corrections in large scale structure modelling. Convolutional kernel weights are shared for kernel pixels equidistant from the centre of a 3D or 2D kernel, associated to the spherical harmonic coefficients $Y^{\ell_k}_m(\theta, \phi)$. Here we make use of these multipole kernels (MPK) in a 2D setting for information capture, embedding the convergence field using a smaller number of neural weights. This choice of embedding is also likely to improve performance in the presence of (white) noise, as noise artefacts are not distributed with the same rotational symmetry as convergence clustering features.
 
 We first embed each log-transformed tomographic bin into six filters corresponding to the $\ell_k=[0,1,2]$ multipole moments for a $7\times7$ kernel per tomographic bin, which for e.g. $N=128$ corresponds to a $0.19\ \rm deg^2$ receptive field. We illustrate a cartoon example of these symmetric kernels for a $3\times 3$ kernel in Figure \ref{fig:neuralnet}. This output is then passed to a nonlinearity and then to another multipole kernel for each input filter, which are then summed along the filter axis at each multipole kernel to yield six output channels. We learn the residual from the first embedding layer $l$ to the next, e.g. $x^{l + 1} = {\texttt{mpk\_layer}}(x^{l} + \texttt{ mpk\_layer}(x^{l}))$. 
 This choice of data embedding layer drastically reduces the number of learnable weights and forces the network to learn physically-symmetric functions of the data in its first layer. {The largest model considered here contains just $6,904$ trainable parameters, which is 0.08\% the footprint of the ResNet18 employed e.g. by \cite{sharma2024comparative, lanzieri2024optimalneuralsummarisationfullfield}. }
\newline

\noindent \textbf{Incept-Stride Tree Network.} The embedded data are then passed to an inception-style network \citep{szegedy2016inceptionv4} with one important difference: instead of varying kernel sizes, we keep the kernel shape fixed to $2\times2$ and vary the \textit{stride} that each layer takes in parallel passes over the data. The objective of this section of the network is to downsample the embedded data by combining information from different scales so that the only features on informative scales are strongly activated and pushed through the learned network to the output summaries using the fewest independent kernel weights  possible.

The data is passed to stride-2, stride-3, and stride-4 downsampling layers followed by a nonlinear activation function and a subsequent stride-1 convolution. The outputs of the stride-3 block are padded periodically in the spatial dimension and concatenated to the output of the stride-2 block, and then passed to another stride-1 convolution. The stride-4 outputs are kept aside until the data has been passed to the next inception block and the data has reached the same spatial resolution. We continue downsampling in this tree-like fashion until the data reach a spatial resolution of $8\times8$. We then mean-pool the features along the spatial axes and pass the resulting flattened filter axis to a final linear layer that outputs the desired additional summaries. Every layer is followed by a new, bijective \texttt{smooth\_leaky} activation function,  
which we found empirically extracted information most consistently across our experiments:
\begin{equation}
    \mathrm{ \texttt{smooth\_leaky}}(x) =
    \begin{cases}
      x, & x \leq -1 \\
      - |x|^3/3, & -1 \leq x < 1 \\
      3x & x > 1.
    \end{cases}
\end{equation}
The intuition here is that unlike natural image data, lensing shear maps are relatively smooth functions, so are best linked to smoother activation functions following convolutions, in contrast to natural images with sharp features like feature borders, for which typical disjoint activations like the \texttt{leaky\_ReLU} were developed \citep{xu2015empirical_activations}.
\newline

\noindent \textbf{Training Setup.} To train the network we split our dataset into equally-sized validation and training sets, with the same $n_s$ number of fiducial and $n_d$ seed-matched derivative simulations. Every epoch a new noise realisation is added onto the noisefree convergence maps and a random rotation is performed. These transformations are seed-matched for each derivative simulation index. We use the \texttt{adam} optimiser with a fixed learning rate of $0.0005$ with gradient clipping at a value of 1.0, and a weight decay penalty of $0.0005$ added to the loss function. These two modifications to the optimisation routine ``smooths'' the loss landscape and prevents the network from overfitting to the training data, respectively. Training is halted when the validation loss stops decreasing significantly for a \texttt{patience} number of epochs.
\newline

\noindent \textbf{Noise Hardening.} All networks are first trained at a low noise level, $A_{\rm noise}=0.125$, after which the noise level is increased in increments of $\Delta A_{\rm noise}=0.05$ for a minimum of 100 epochs subject to a patience setting of 75 epochs at each setting. This can be thought of as ``domain-transfer'' learning on-the-fly. Slowly increasing the noise allows physical features (e.g. convergence patterns) to be embedded early in training, such that the network outputs are already concentrated on the informative distribution of the data when the shape noise increases.

\begin{figure}
    \centering
    \includegraphics[width=0.75\columnwidth]{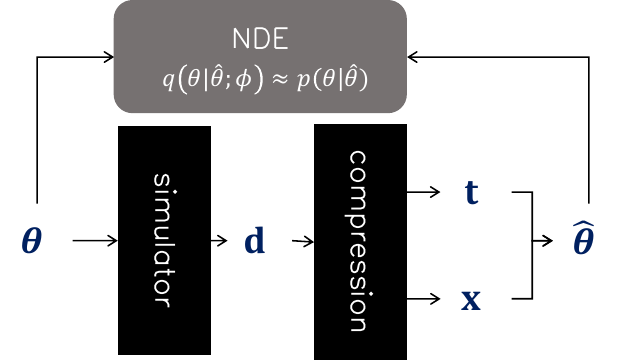}
    \caption{Cartoon of density estimation scheme with fixed compression (network or MOPED). Parameters $\parvec$ are drawn from a prior and MLE estimates $\hat{\parvec}$ are produced from data $\textbf{d}$ for either (fixed) compression method using Eq. \ref{eq:mle-from-fisher} and fed to a MAF neural density estimator for the amortised posterior distribution, trained under the loss in Eq. \ref{eq:nde_loss}.}
    \label{fig:sbi-cartoon}
\end{figure}

\subsection{Neural Density Estimation}\label{sec:nde}
To measure the information capture in both MOPED and network summaries we employ a neural posterior estimation scheme to parameterise the amortised summary-parameter posterior $p(\parvec | \textbf{y})$, where $\textbf{y}(\textbf{d})$ is either the MOPED summary or updated summary set of MLE parameter estimates using Eq. \ref{eq:mle-from-fisher}. We employ an identical ensemble of masked autoregressive flows  \citep[MAFs;][]{papa_mafs2017, pydelfiAlsing_2019} for each set of summaries using the LtU-ILI codebase \citep{ho2024ltuili}. We opt for networks with 50 hidden units and 12 transformations. We chose this high level of complexity such that the posterior density parameterisations in all cases were sufficiently descriptive. A unique aspect of our network training scheme is that a joint parameter-data prior distribution can be chosen after learning the network and MOPED compression, displayed as a cartoon in Figure \ref{fig:sbi-cartoon}. We generated 5000 simulations for each of two wide uniform priors in the $S_8$ formalism: $p^{(1)}(\Omega_m, S_8) = \mathcal{U}[[0.15, 0.35] \times [0.7, 1.52]]$ and $p^{(2)}(\Omega_m, S_8) = \mathcal{U}[[0.15, 0.35] \times [0.5, 1.0]]$. We opt for the smaller prior in cases where the $3\sigma$ Fisher posterior estimate for the observable considered falls within the support of $p^{(2)}$.

\begin{figure}[htpb!]
    \centering
    \includegraphics[width=\columnwidth]{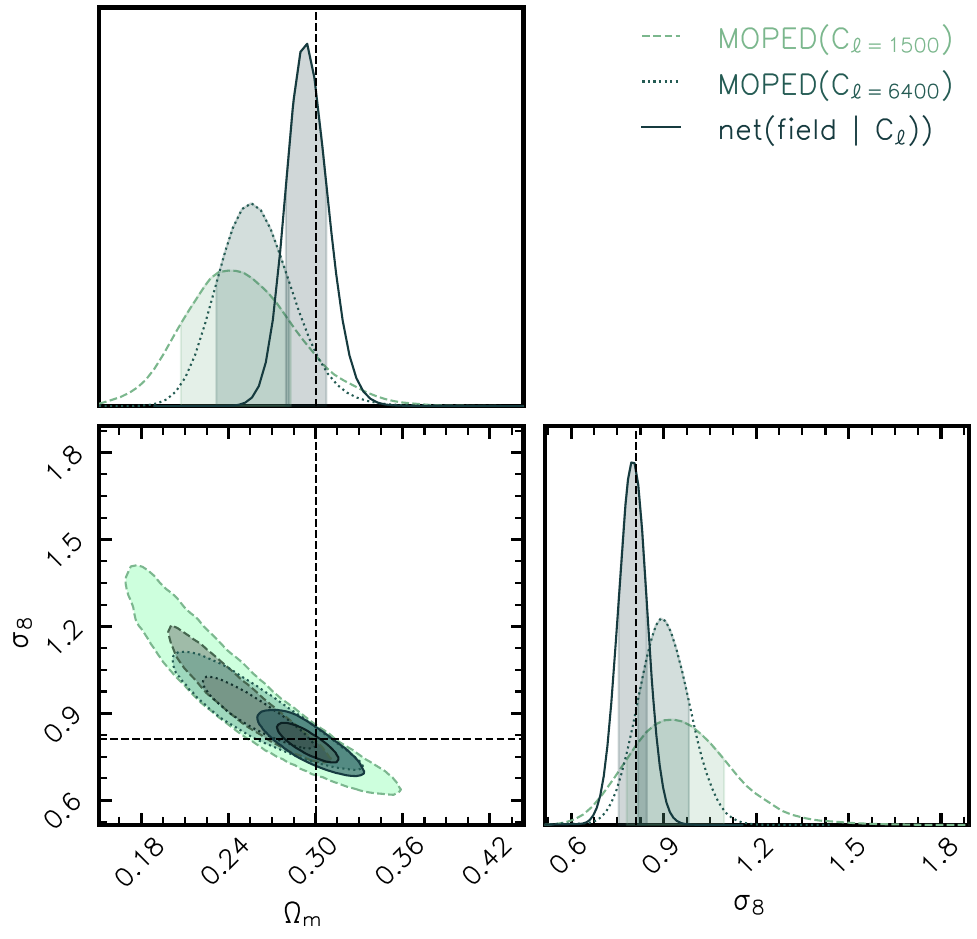}
    \caption{Using information-update network summaries (green) drastically improves $\Omega_m - \sigma_8$ constraints beyond MOPED $C_\ell$ summaries in a low-noise setting. We compare the posteriors obtained from a KiDS-like survey truncation at $\ell_{\rm cut}=1500$ (blue) to the constraints from all available modes $\ell_{\rm cut}=6400$ at the given resolution (green). The network's additional summaries (dark green) is able to improve information extraction by a factor of 5 beyond the $\ell_{\rm cut}=6400$ and a factor of 8 above $\ell_{\rm cut} = 1500$.}
    \label{fig:threeblobs}
\end{figure}

\begin{figure}[htpb!]
    \centering
    \includegraphics[width=\columnwidth]{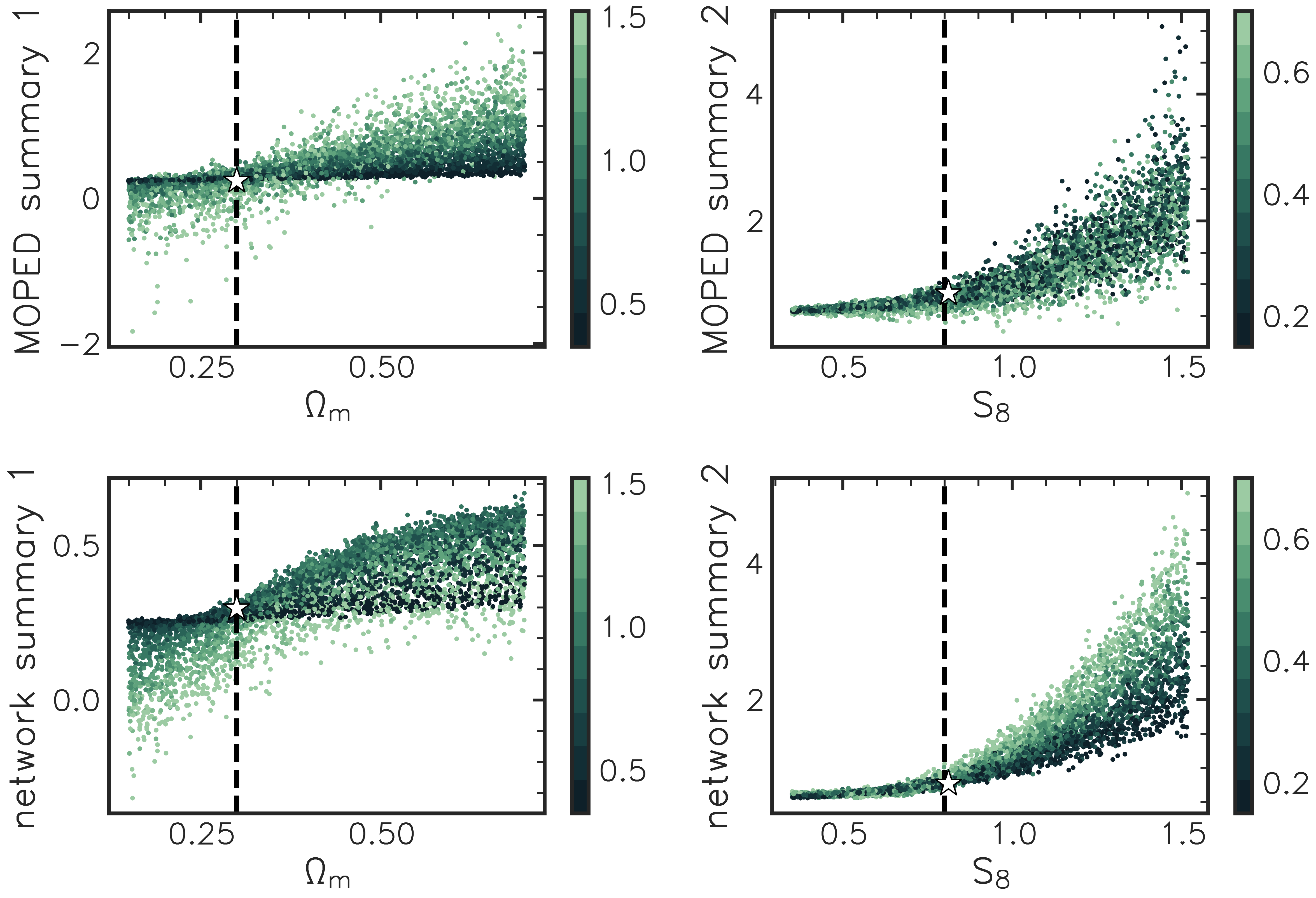}
    \caption{Information-update network (bottom) makes simulations more distinguishable in summary space than $C_\ell$ compression (top). Points in parameter-summary space are coloured by the opposite parameter's value. The network finds patterns that separate these summaries in a complementary fashion even away from the fiducial point $(\Omega_m, S_8 )= (0.3, 0.8)$. We display a 3D view of this four-dimensional space in Figure \ref{fig:3dsummaries}.}
    \label{fig:summary-scatter}
\end{figure}

\begin{figure*}[htp!]
    \centering
    \includegraphics[width=0.9\textwidth]{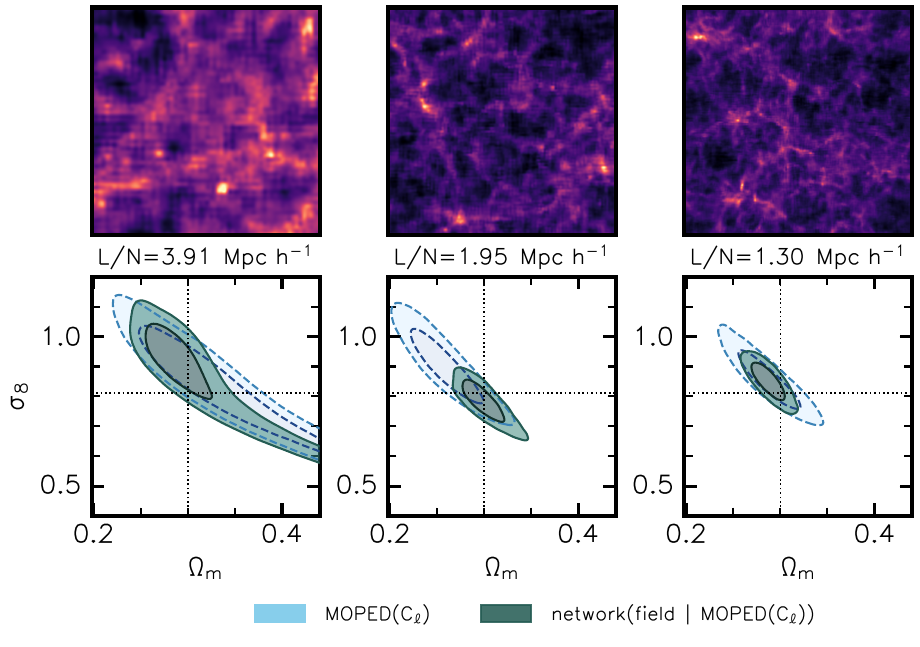}
    \caption{Computing additional complementary summaries from the convergence field improves parameter constraints (green) over the two-point information (blue) as the field becomes more nonlinear in the low-noise regime. For $N=64$ fields the information gain above the $C_\ell$ constraints is modest, but improves as more nonlinear scales are included at the level of the field as resolution increases.}
    \label{fig:resolution-comparison}
\end{figure*}

\section{Results}

\subsection{Low-noise Regime}\label{sec:results}
We first investigate the information extraction as a function of dark matter simulation resolution with a small amount of shape noise, and compare information extraction to two-point statistics at different scales.  We construct particle mesh simulations with varying numbers of pixels $N_x = N_y \in [64, 128, 192]$. Intuitively we expect more information beyond the two-point statistic to be found at higher resolutions, which can be interpreted as the descriptiveness of the underlying gravity model.
\newline
\newline
\noindent \textbf{Scale Cutoff.} We first explored the effect of a scale cutoff at resolution $N=128$ for the $C_\ell$ summaries with a low noise setting. We construct MOPED summaries from $C_\ell$s subject to a Stage-III survey-like cut at $\ell_{\rm cut}=1500$, as well as summaries from all available $\ell$ modes at the given resolution. The highly-compressed MOPED summaries give almost identical posteriors to using the full set of $C_\ell$ values as the data vector. The network is tasked with finding complementary summaries in the $\ell_{\rm cut} = \ell_{\rm max}$ case. We display the constraints obtained on the same target simulation in Figure \ref{fig:threeblobs} and in Table \ref{tab:results}. The network extracts up to 5 times more information than the two-point function in a low-noise setting with all modes and 8.3 times more information in high-noise settings, as measured by the determinant of the Fisher matrix.

\begin{table}[ht!]
\begin{center}
\adjustbox{width=\columnwidth}{%
 \begin{tabular}{l l r r r}
 \toprule
 & resolution  &  {$ H(C_\ell)$} & {$ H({\rm net})$} & ratio \\
 \hline
    & $N=64$   & $6.9$ & ${7.4}$ & $\boldsymbol{2.9}$  \\
    low noise & $N=128$  & $6.9$ & ${7.7  }$ & $\boldsymbol{5.0}$ \\
    & $N=192$  & $7.6$ & ${8.5}$ & $\boldsymbol{5.1}$ \\
\hline
    high noise & $N=128$  & $5.3$ & ${6.0}$ & $\boldsymbol{4.5}$ \\
    & $N=192$  & $5.2$ & ${6.3}$ & $\boldsymbol{8.3}$ \\

 \toprule
\end{tabular}
}
    \caption{
    {\textnormal{Summary of parameter Shannon information (H = $\frac{1}{2} \ln \det F$) from MOPED and information-update networks for low noise ($n_{\rm gal} = 1900$) and high noise ($n_{\rm gal} = 83$) scenarios. The ratios of Fisher determinants are shown in the last column.} For the noisy $N=192$ case we optimise networks against the binned $C_\ell$ vectors as opposed to the MOPED summaries.}
    }
\label{tab:results}
\end{center}
\end{table}

\noindent \textbf{Summary Scatter.} 
The information-update loss scheme asks the network to use data features such that output summaries complement existing $C_\ell$-based summaries. We can interpret the statistics learned by the network by visualising a summary scatter over the suite of prior simulations. Figure \ref{fig:summary-scatter} shows the network and MOPED outputs versus true parameter, coloured by the opposite parameter's value for summaries used to generate the network and $\ell_{\rm max}=6400$ posteriors in Figure \ref{fig:threeblobs}. Remarkably, even though the network is only trained at the fiducial cosmology (dashed vertical line in each plane), the information-update loss allows the network to find useful features with which to distinguish parameters {in a smoother summary space (less scatter) than the MOPED compression. This increased structure is then harnessed by the density estimation scheme to provide tighter parameter constraints than MOPED. The complementary nature of the information from the network-updated statistics to the original statistics decreases away from the fiducial point in both dimensions, but does so smoothly, i.e. the information about the parameters coming from the four statistics is mixed away from the fiducial point.
\newline

\noindent \textbf{Resolution Dependence.} We next compare constraints as a function of PM simulation resolution, which effectively controls the nonlinearity of the dark matter gravity solver. Here we wish to measure the parameter information gain that using a nonlinear network to probe nonlinear scales adds to the power spectrum. We generate three suites of fiducial and finite-difference convergence maps to learn the compression with $A_{\rm noise} = 0.125$. Figure \ref{fig:resolution-comparison} shows constraints at each resolution. More non-Gaussian information is extracted for the higher resolution simulations, indicated by the increase in network constraining power over the $C_\ell$s, since these simulations probe smaller, more nonlinear scales accessed by the network. We report our network and MOPED Fisher constraints in Table \ref{tab:results}. In this low-noise setting we observe an information increase of a factor of 2.9 for $N=64$ and a factors of 4-5 for $N=[128,192]$ high-resolution simulations, aligning with our intuition.

\subsection{High-noise Regime}\label{sec:highnoise}
The information-update formalism displays promising results in the presence of increased systematics such as galaxy shape noise. Here we start with a network trained on the lowest noise setting and slowly increase the noise amplitude (equivalently decreasing the galaxy density). The network is able to increase its relative performance against the two-point statistic as we increase resolution (Fig. \ref{fig:noise-resolution-comparison}) and shape noise (Fig. \ref{fig:noise-comparison}). Figure \ref{fig:noise-resolution-comparison} shows that with increased simulation resolution the network has access to more nonlinear scales and can compensate for the shot noise that dominates the $C_\ell$ calculation at high $\ell$ values. For a fixed resolution ($N=192$), the additional summaries found by the network appear robust to noise; keeping the parameter constraints consistent as increased shape noise pushes the $C_\ell$ constraints towards the prior edge in $\Omega_m$. This is especially promising for higher noise cases for galaxy density $n_{\rm gal} < 50\ {\rm arcmin}^{-2}$, as this falls between the capabilities of Euclid \citep{Euclidcollab2021} and Roman \citep{spergel_roman} telescopes. \newline
\newline
\noindent \textbf{Optimising With the Binned Power Spectrum.} For our high-resolution simulations we also explored optimising the information-update formalism (Eq. \ref{eq:matrixupdate}) with respect to the full binned $C_\ell$ vector. This ``stretches'' the off-diagonals $\textbf{u}$ of the full summary covariance (Eq. \ref{eq:fullcov}), such that the IUF forces the network to respond to explicit fluctuations in these $C_\ell$ bins with respect to noise. Here we posit that the network will be able to find summaries that complement fluctuations at different $\ell$ scales more efficiently. We find that indeed this choice of optimisation allows the network to extract $8.3$ times more information than the two point function in the noisiest ($n_{\rm gal}=30$) setting, compared to a $5.6$ times improvement when optimising against the MOPED-compressed summaries. We visualise the joint covariance of learned and binned $C_\ell$ summaries (Eq. \ref{eq:fullcov}) in Appendix \ref{app:cov}.

\begin{figure}
    \centering
    \includegraphics[width=\columnwidth]{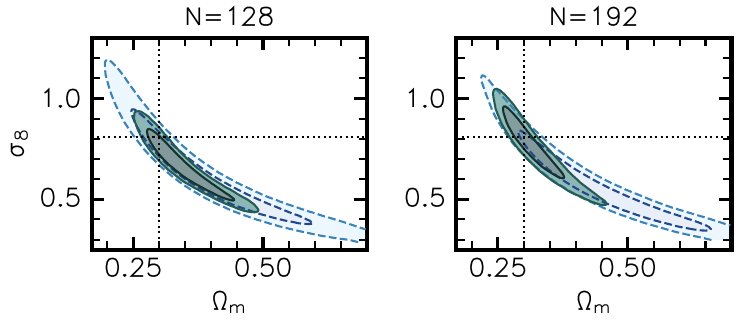}
    \caption{Additional neural summaries (green) are robust to shape noise $(n_{\rm gal} = 120)$ at different resolutions. Constraints from $C_\ell$-only summaries (blue) suffer from the increased noise due to shot-noise contributions to the high-$\ell$ bins.}
    \label{fig:noise-resolution-comparison}
\end{figure}

\begin{figure}
    \centering
    \includegraphics[width=\columnwidth]{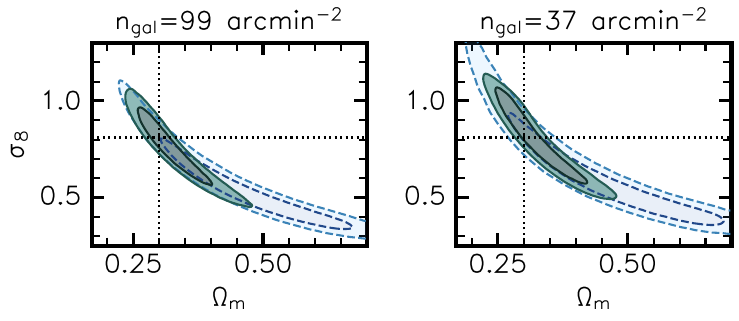}
    \caption{Neural summaries (green) are robust to increased shape noise, controlled by the galaxy density parameter $n_{\rm gal}$. Increased shot noise at small scales inhibits angular $C_\ell$ constraints (blue). Here we display an inference on the same $N=192$ resolution simulation subject to increased shape noise. We optimise the network using the binned $C_\ell$ vectors.}
    \label{fig:noise-comparison}
\end{figure}

\section{Discussion \& Conclusions}
In this paper, we present an implicit inference technique to extract neural summary statistics from field-level data, specifically weak lensing maps,  that are designed to match or increase automatically the Fisher information about the cosmological parameters over a set of pre-defined summaries, typically traditional two-point statistics. We apply this method to find summary statistics from tomographic convergence maps that explicitly complement the angular power spectrum estimates. This powerful hybrid mixture of physics-based and neural network derived summary statistics is  guaranteed to improve the two-point parameter constraints and allows for networks with small physics-informed architecture to achieve similar results to larger regression networks. We demonstrated that this approach extracts between a factor 3 and 8 more information than the angular power spectrum, as measured by the determinant of the Fisher matrix.  For weak lensing, the main gain is a substantial reduction in the credible region for $\Omega_m$, with a smaller improvement in the $S_8$ error.
\newline

\noindent \textbf{What is Being Learned.} Neural networks should be thought of as a mapping to a manifold whose shape is controlled by the network architecture. Here we restrict our manifold to be translationally-invariant (convolutional) and impose that it resides in the space parameterised by weight-sharing multipole kernels. Furthermore, we optimise this lightweight architecture to find summaries that can only complement estimates of the angular power spectrum. This effective truncation of function space also contributes to the interpretability of the network, as it guarantees that the function learnt is not the power spectrum, or Gaussianly-compressed summaries of it. \newline

Other studies have previously combined power spectrum and network summaries in weak lensing analyses. \citet{jeffrey2024dark} for example feed in both sets of independently-obtained summaries into an NDE for posterior estimation to obtain a $\sim 2\times$ improvement in information extraction in $\Omega_m - S_8$. Here we show that coordinating the field-level network optimisation with an existing summary can give us even more efficient extraction.

\noindent \textbf{Asymptotic Optimality.} Unlike regression networks, which are trained over the entire prior distribution, our network learns the compression locally around a single point. This prevents the network from learning features of the convergence map that appear in unlikely points in parameter space. Once the compression from data to network summaries is learned, simulations over a prior $p(\parvec, \dat)$ can be one-shot compressed before being fed to an NDE posterior.
\newline

The hybrid summary formalism presented in this work is not limited to weak lensing data, and it can be generalised to any dataset to identify the features from which the information captured by large neural networks comes. This technique might also reduce the need for large convolutional networks to learn the large-scale correlations in larger dark matter and galaxy simulations \citep{lemos2023simbigfieldlevelsimulationbasedinference}. In future work, we will apply this formalism to find the summary that complements the information from more than one pre-determined summary statistic, such as angular power spectrum and peak count summaries. This has the potential to improve the cosmology constraints from implicit likelihood analyses of weak lensing such as \citet{fluri_kids} and \citet{jeffrey2024dark}.

\section{Code Availability}
The code for this analysis will be made available at \url{https://github.com/tlmakinen/hybridStatsWL}. All custom networks and simulation tools were written in \texttt{Jax} \citep{jax2018github} and \texttt{flax} \citep{flax2020github} and were run on a single NVIDIA v100 32Gb GPU. Posterior density estimation was performed locally on a laptop CPU using the LtU-ILI code \citep{ho2024ltuili}. 

\section{Acknowledgements}
TLM acknowledges the Imperial College London President's Scholarship fund for support of this study, and thanks Niall Jeffrey, Justin Alsing, David Spergel, and Maximilian von Wietersheim-Kramata for insightful discussions. NP is supported by the Beecroft Trust. 
BDW. acknowledges support by the ANR BIG4 project, grant ANR-16-CE23-0002 of the French Agence Nationale de la Recherche;  and  the Labex ILP (reference ANR-10-LABX-63) part of the Idex SUPER, and received financial state aid managed by the Agence Nationale de la Recherche, as part of the programme Investissements d'avenir under the reference ANR-11-IDEX-0004-02. TLM acknowledges helpful conversations facilitated by the \href{https://www.learning-the-universe.org/}{Learning the Universe Collaboration}.
The Flatiron Institute is supported by the Simons Foundation.

\bibliography{mybib}
\bibliographystyle{aasjournal}

\myemptypage
\appendix 


\section{Posterior Coverage Tests}\label{app:coverage}
One of the distinct advantages of SBI neural density estimation is the immediate availability of coverage tests. In this work we trained an estimator for the posterior distribution given some point-estimates for the parameters via MOPED or the hybrid-summary network: $p(\parvec | \hat{\parvec})$.
This density estimator is an \textit{amortised} posterior, meaning the posterior density for any given summaries $\hat{\parvec}$ is immediately available without having to do MCMC sampling with a likelihood. We can then do repeated mock data parameter inference over the prior using this posterior density, and calculate how many true parameter vlaues from the credible intervals match the expected fraction, forming a posterior ``coverage'' test. We display one such test in Figure \ref{fig:tarp-coverage} making use of the \texttt{TARP} coverage test framework presented in \cite{lemos2023samplingbasedaccuracytestingposterior}. \newline

\begin{figure}[htp!]
    \centering
    \includegraphics[width=0.9\columnwidth]{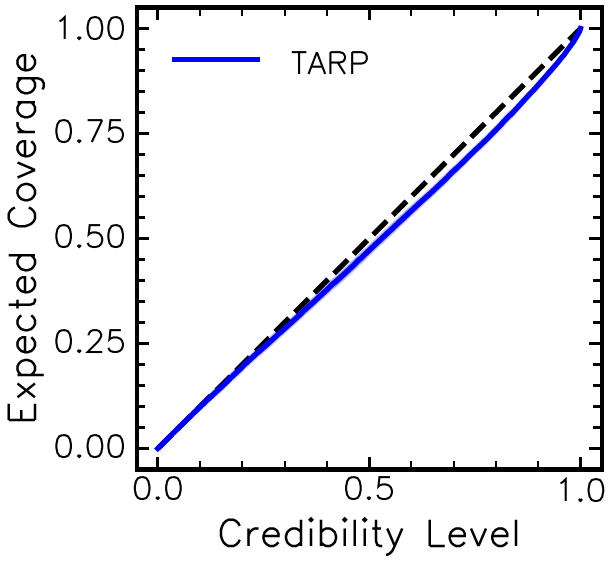}
    \caption{Example coverage test result for low-noise inference with $N=192$ resolution (rightmost panel in Fig. \ref{fig:resolution-comparison}) using \texttt{TARP} \citep{lemos2023samplingbasedaccuracytestingposterior}. Using our amortised parameter-summary posterior $p(\parvec, \hat{\parvec})$, we can do repeated mock data parameter inference over the prior, and measure which fraction of true values from the appropriate credible intervals matches the expected fraction. The blue line traces 100 ``distances to random points" (DRP), which is accelerated using the \texttt{TARP} framework within LtU-ILI \citep{ho2024ltuili}. The DRP line (blue) traces the truth line (dashed), indicating a successful test.}
    \label{fig:tarp-coverage}
\end{figure}

\begin{figure*}[htp!]
    \centering
    \includegraphics[width=\textwidth]{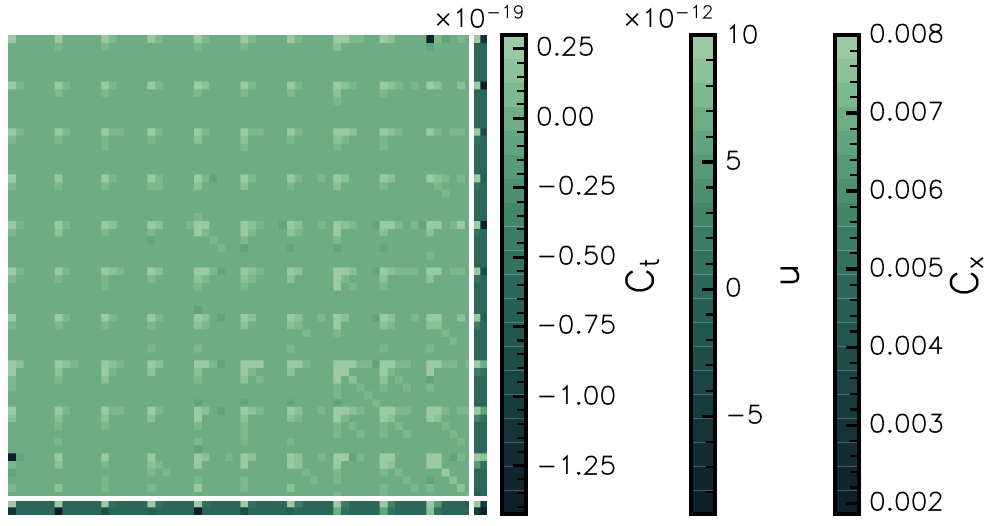}
    \caption{Example joint $C_\ell$-network summary covariance visualisation (Eq. \ref{eq:fullcov_vector}) for a network optimised against the binned power spectrum. We separate the $60\times 60$ $C_\ell$ covariance structure (upper left corner) from network summaries (lower right corner) with the set of intersecting white lines. The learned network summaries exhibit a non-zero correlation structure with the $\ell$ bins, illustrated by the $\textbf{u}$ off-diagonal matrix vectors on the bottom and right-hand edges. Here it is obvious that only one of the two network summaries is highly correlated with the binned power spectrum modes.}
    \label{fig:covexample}
\end{figure*}

\section{Learned Covariance Matrix Visualisation}\label{app:cov}
In Figure \ref{fig:covexample} we illustrate the full joint covariance,
\begin{equation}\label{eq:fullcov_vector}
    \textbf{C} = \begin{pmatrix}
    \textbf{C}_t & \textbf{u}\\
    \textbf{u}^T & \textbf{C}_x
\end{pmatrix},
\end{equation}
of the binned tomographic $C_\ell$ statistic and two learned network summaries, clearly separated by the white intersecting lines. We plot each component of this structure with a separate colourbar. The cross-correlation row-matrices $\textbf{u}$ indicate that the learned summaries exhibit non-trivial correlation structure with the binned $\ell$-modes, which contributes to information capture according to the hybrid statistics formalism.

\section{Summary Scatter}\label{app:summaryscatter}
\begin{figure*}[htp!]
    \centering
    \includegraphics[width=0.85\textwidth]{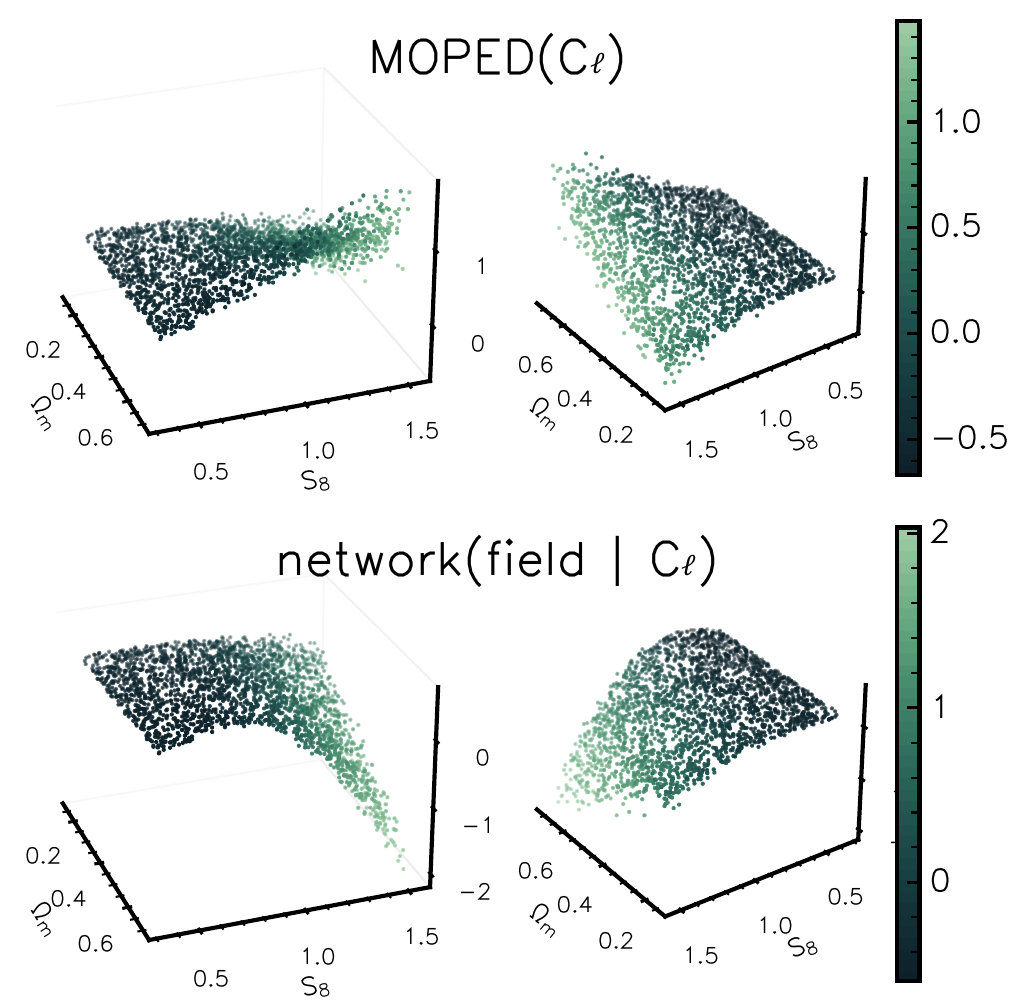}
    \caption{Computing additional information from the data endows the network summaries (bottom row) with more structure as a function of parameters $(\Omega_m, S_8)$ over a prior than MOPED two-point summaries (top row). Summaries for each method $\hat{\Omega}_m$ and $\hat{S}_8$ are indicated by $z$-direction and colourbar, respectively. It is highly visible via the increased structure in joint space that the network is able to capture information from the smaller scales it has access to. More scatter in $z$ or colour at a particular parameter value in the MOPED summaries indicates a less informative compression of the simulated convergence data to from power spectrum summaries. }
    \label{fig:3dsummaries}
\end{figure*}
Here we display a three-dimensional view of the four-dimensional joint distribution of compressed summaries and parameters $p(\parvec, \hat{\parvec})$ obtained from MOPED and network compressions in a low-noise setting. The information update formalism tells the convolutional network during optimisation to make use of the nonlinear information on the smaller (pixel-level) scales that it has access to in a way that is complementary to the power spectrum. Although optimised at a fiducial point, the mapping learned is smooth as a function of data $\textbf{d}(\parvec)$ away from the training point, resulting in more structure in the four-dimensional joint distribution space $p(\parvec, \hat{\parvec})$ than MOPED, allowing the summaries ($z$-axis and colour) to respond more smoothly and rapidly as a function of parameters $(x,y)=(\Omega_m, S_8)$. This smoother joint distribution surface can then be harnessed by the NDE scheme to produce tighter posteriors in an amortised fashion.

\end{document}